\documentclass{article}
\usepackage{spconf,amsmath,graphicx,hyperref}
\usepackage{booktabs}
\usepackage{subcaption}
\usepackage{xcolor}

\title{Toward Faithful Explanations in Acoustic Anomaly Detection}
\name{\smash{\parbox{0.80\linewidth}{%
\begin{center}
Maab Elrashid$^{1,2,3,4}$, Anthony Desch\^enes$^{3,4}$, Cem Subakan$^{1,2,4}$, 
Mirco Ravanelli$^{1,2}$, R\'emi Georges$^{3,4}$, Michael Morin$^{3,4}$
\end{center}}}}
\address{%
$^1$ Mila-Quebec AI Institute, QC, Canada, $^2$ Concordia University, QC, Canada\\[1ex]
$^3$ FORAC Research Consortium, 
$^4$ Université Laval, QC, Canada
}
\begin{document}
\ninept
\maketitle
\newcommand{\cem}[1]{\textbf{Cem: #1}}
\begin{abstract}

Interpretability is essential for user trust in real-world anomaly detection applications. However, deep learning models, despite their strong performance, often lack transparency. In this work, we study the interpretability of autoencoder-based models for audio anomaly detection, by comparing a standard autoencoder (AE) with a mask autoencoder (MAE) in terms of detection performance and interpretability. We applied several attribution methods, including error maps, saliency maps, SmoothGrad, Integrated Gradients, GradSHAP, and Grad-CAM. Although MAE shows a slightly lower detection, it consistently provides more faithful and temporally precise explanations, suggesting a better alignment with true anomalies. To assess the relevance of the regions highlighted by the explanation method, we propose a perturbation-based faithfulness metric that replaces them with their reconstructions to simulate normal input. Our findings, based on experiments in a real industrial scenario, highlight the importance of incorporating interpretability into anomaly detection pipelines and show that masked training improves explanation quality without compromising performance.

\end{abstract}
\section{Introduction}
\label{sec:intro}

Anomaly detection plays a critical role in industrial monitoring, where early fault identification is essential~\cite{deschenes2025planing}. For example, in wood manufacturing, equipment like planers relies on manual calibration, with limited real-time feedback on machine condition or product quality. Acoustic monitoring with low-cost microphones offers a promising alternative for anomaly detection, but remains challenging in noisy industrial environments where subtle acoustic cues can be easily masked~\cite{deschenes2025planing, BUEHLMANN2002197}.
Recent deep learning approaches, particularly convolutional autoencoders with skip connections and transformer-based architectures, show superior performance in acoustic
anomaly detection in this context~\cite{deschenes2025planing}. However, their black-box nature limits trust in safety-critical environments, as models can exploit spurious features that correlate with anomalies but are not truly relevant~\cite{li2023survey,lapuschkin2019unmasking}. Interpretability avoids this by designing accurate models with correct explanations that enhance transparency and user trust \cite{li2023survey}.

In this work, we investigate the interpretability of anomaly detection using autoencoders
on real industrial data from a wood planer~\cite{deschenes2025planing}. We implement masked autoencoder (MAE) training~\cite{he2022masked} to improve feature learning. We apply several post-hoc explanation techniques, such as error maps and gradient-based methods~\cite{li2023survey}, to analyze influential input regions. To validate the explanations, we match the anomaly intervals marked by experts and propose two metrics: an F-score for temporal overlap and a faithfulness score to assess the impact of replacing highlighted regions on the model output\footnote{The annotated dataset and code are available at https://github.com/Maab-Nimir/Faithful-Explanations-in-Acoustic-Anomaly-Detection}.

In summary, our contributions include: (1) An interpretability study of gradient-based methods applied to autoencoders for acoustic anomaly detection; (2) adapting MAE-based training to improve anomaly localization; and (3) an evaluation protocol based on expert-annotated time segments and quantitative metrics that assess explanation relevance and faithfulness.

\section{Related Work}
\label{sec:literature}

Autoencoders (AEs) are widely used for unsupervised anomaly detection (AD), due to their ability to learn compact representations and identify anomalies by reconstruction error~\cite{Koizumi2020DescriptionAD, marchi2015novel, zimmerer2019unsupervised}. 
However, AEs often struggle with precise anomaly localization. 
They can reconstruct anomalous instances, reducing error-based anomaly signals~\cite{bouman2025autoencoders}. 
Additionally, they can react on irrelevant correlations, making interpretability important in critical applications~\cite{li2023survey}.

To address these limitations, several works have applied post-hoc interpretability methods to AD using AEs, such as SHAP~\cite{roshan2021utilizing, antwarg2019explaining}, saliency maps, integrated gradients and SmoothGrad, 
which highlight input regions influence the anomaly score~\cite{li2023survey}. 
Yet, in unsupervised settings,
such methods often lack clarity and robustness~\cite{tritscher2023feature}.

Beyond post-hoc methods, newer AE frameworks have been explored to inherently improve interpretability. Masked Autoencoders (MAEs) were originally proposed for self-supervised learning in vision,
to learn semantically meaningful features by reconstructing masked inputs~\cite{he2022masked}. This strategy has been extended to AD in computer vision~\cite{huang2022self0supervised}, medical imaging~\cite{georgescu2023masked} and time series~\cite{lee2024explainable}. However, to the best of our knowledge, MAE-based methods, especially those tailored for spectrogram‑based audio anomaly detection aimed at interpretability, have not yet been explored.

Furthermore, recent work emphasizes the importance of evaluating the explanation quality~\cite{li2023survey, chan2022comparative}. Metrics such as the F-score assess plausibility against human-annotated anomalies~\cite{holly2022autoencoder}, while Faithfulness measures how well the model explanations align with the model decision~\cite{vsimic2025comprehensive}.

Our work builds on this foundation by applying MAE to industrial audio anomaly detection, with emphasis on interpretability rather than detection accuracy. We adapt the MAE framework to reconstruct masked spectrogram regions and derive explanations via error maps and gradient-based methods. Additionally, we introduce time-localized F-score and faithfulness metrics using expert annotations to quantify how well explanations align with human perception and model's decision making.

\section{Methodology}
\label{sec:method}


\subsection{Real Planing Mill Dataset}
We use the publicly available dataset from~\cite{deschenes2025planing}, which comprises 7,562 mono 10-second recordings of an industrial wood planer, sampled at 20 kHz (\raisebox{0.6ex}{\texttildelow}21 hours in total). Each audio clip is converted into an $80 \times 401$ mel spectrogram with a frame size of 50 ms, hop length of 25 ms, and 80 mel frequency bins.

The training set contains only normal recordings captured after maintenance and 10\% of it is used for validation.
The test set spans two days of planer operation and includes normal and anomalous samples (broken, stuck, and unevenly thick boards). Table~\ref{tab:dataset_stats} summarizes the data.

\begin{table}[t!]
\centering
    \caption{Dataset composition and anomaly types.}
    \footnotesize
\begin{tabular}{l c c c}
        \toprule
         & \textbf{Training set (normal)} & \multicolumn{2}{c}{\textbf{Test set (normal + anomalous)}} \\
                \midrule
        \textbf{Count} & 4,327 & \multicolumn{2}{c}{3,235}\\
        \midrule
        & \multicolumn{3}{c}{\textbf{Anomaly types distribution (test set)}} \\ \cmidrule(lr){2-4}
        & Broken Board & Stuck Board &  Uneven/Thick Board \\
        \midrule
        \textbf{Count}
          & 4 
          & 29 
        & 72 \\
        \bottomrule
    \end{tabular}
    \label{tab:dataset_stats}
\end{table}

\subsection{Anomaly Detection Models}

\subsubsection{Autoencoder}
We adopt the Skip-CAE-Transformer from~\cite{deschenes2025planing}; a reconstruction-based model that combines a convolutional autoencoder with skip connections and transformer blocks~\cite{vaswani2017attention} for anomaly detection. 
The encoder extracts hierarchical representations from the input spectrogram using convolution, batch normalization, and pooling, followed by a transformer encoder. Skip connections are added and passed to the decoder, which mirrors the encoder structure and includes a transformer decoder for improved reconstruction. 

The model is trained to minimize the Mean Squared Error (MSE) between the input and the reconstructed output.

\subsubsection{Masked Autoencoder}

Inspired by~\cite{he2022masked}, we apply an adapted masked autoencoder (MAE) training strategy to our model. The model receives a randomly masked input $X$ and learns to reconstruct only the masked regions, using the loss:
$$\mathcal{L}_{\text{MAE}} = \frac{1}{\sum\limits_{i,j} M_{ij}} \sum_{i,j} M_{i,j} (X_{i,j} - \hat{X}_{i,j})^2$$
where, $M$ is the binary mask and $\hat{X}$ is the reconstructed output.

This partial reconstruction objective forces the model to rely on contextual cues, which encourages meaningful feature learning. 


We conducted an ablation study to select the mask ratio and patch size. Table~\ref{tab:mae_ablation} shows that 30\% masking with $4 \times 4$ patches yields the best AUC (Area Under the ROC Curve), while extreme ratios or larger patches degrade performance~\cite{kong2023understanding}. Additionally, to ensure divisibility by patch size, the final time frame is removed, reducing the input to 400 frames with a negligible effect.

\begin{table}[t!]
    \centering
    \caption{AUC for MAE with varying mask ratios and patch sizes.}
    \footnotesize
    \begin{tabular}{ccc}
        \toprule
        \textbf{Mask Ratio (\%)} & \textbf{AUC $4 \times 4$ Patch Size} & \textbf{AUC $16 \times 16$ Patch Size} \\
        \midrule
        90  & 0.842 & 0.808 \\
        75  & 0.864 & 0.710 \\
        50  & 0.842 & 0.710 \\
        30  & \textbf{0.902} & 0.636 \\
        15  & 0.821 & 0.569 \\
        \bottomrule
    \end{tabular}
    \label{tab:mae_ablation}
\end{table}

At inference, the model receives the full (unmasked) spectrogram and outputs a full reconstruction. Anomaly scores and attribution maps are computed using the mean squared error between input and output~\cite{georgescu2023masked}.

\subsection{Interpretability Methods}

To analyze the model and localize anomalies, we apply several post-hoc interpretability methods using the Captum library\footnote{https://captum.ai/}. For gradient-based methods, we use the reconstruction mean squared error (MSE) as the scalar output for backpropagation~\cite{li2023survey}, generating 2D attribution maps over the input spectrogram, highlighting the regions that contribute the most to the reconstruction error.

We apply pixel-wise error map~\cite{zimmerer2019unsupervised}, saliency-based techniques including Saliency Maps~\cite{simonyan2013deep}, Integrated Gradients~\cite{sundararajan2017axiomatic}, and SmoothGrad~\cite{smilkov2017smoothgrad}, as well as GradShap~\cite{lundberg2017unified}, which approximates SHAP values through gradient integration over randomized baselines. Furthermore, we use Grad-CAM~\cite{selvaraju2017grad} to highlight coarse regions by propagating gradients to the final convolutional layer of the encoder.


\subsection{Interpretability Evaluation}

\subsubsection{Human Annotations}
\begin{figure}[t!]
    \centering
    \includegraphics[width=0.7\linewidth]{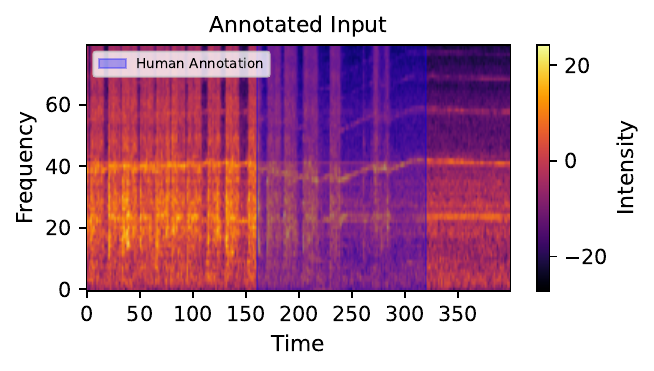}
    \caption{Example of a human annotation for a broken board. The blue shaded region indicates the labeled anomaly interval.}
    \label{fig:annotation_example}
\end{figure}

To assess the alignment between model explanations and human perception, 46 anomalous test recordings were annotated based on expert listening and spectrogram inspection, covering the 3 types of anomalies as in Table~\ref{tab:annotated_samples}. Each annotation consists of one or more clearly audible anomalous intervals of at least one second (40 frames). Recordings with unclear acoustic signatures were excluded for quality. Figure~\ref{fig:annotation_example} shows an example of annotation for a broken board.

\begin{table}[t!]
    \centering
    \caption{Anomaly types distribution in the annotated sample set.}
    \footnotesize
    \begin{tabular}{lccc}
        \toprule
         & \textbf{Broken Board} & \textbf{Stuck Board} &  \textbf{Uneven/Thick Board} \\
        \midrule
        \textbf{Count}
          & 2 
          & 10 
        & 34 \\
        \bottomrule
    \end{tabular}
    \label{tab:annotated_samples}
\end{table}

\subsubsection{Evaluation Metrics}
We use two complementary metrics to evaluate explanation quality: the F-score and Faithfulness.

To compute the \emph{F-score} \cite{powers2020evaluation0}, each 2D attribution map is collapsed into a 1D temporal signal by sum over frequency and normalization. Peaks are identified using high-percentile thresholding (e.g., 98th) and compared to the annotated 1-second intervals~\cite{gungor2024robust}. A true positive (TP) occurs if any peak falls within an annotated second. A false negative (FN) occurs when no peak is detected within an annotated second. A false positive (FP) is when a peak appears in an unannotated second \cite{el2023multivariate}.

\emph{Faithfulness} evaluates whether regions highlighted by an attribution map meaningfully affect the reconstruction error~\cite{chan2022comparative}. 
We propose to estimate how the model would behave if the highlighted input region were instead ``normal''. 
 
Specifically, we replace the anomalous segments detected in the input $X_1$ with their reconstructed counterparts by the model $\hat{X_1}$, resulting in a modified input $X_2$ and reconstruction $\hat{X_2}$. 
$$X_2 = M\cdot\hat{X_1} + (1-M)\cdot X_1$$
where, $M$ is a binary mask over detected regions. We define the Faithfulness Score $FF$ as:
$$
FF = \max\left(1 - \frac{\text{Error}(\hat{X}_2, X_2)}{\text{Error}(\hat{X}_1, X_1)},\ 0\right)
$$
Such that, if the replaced regions are irrelevant to the model’s decision, the reconstruction error changes little, resulting in a low $FF$ score. Conversely, a higher $FF$ means masking those regions significantly affects the error, indicating the explanation reliably highlights areas important to the model’s prediction.

We experimented with frame-based and segment-based replacement strategies. 
In \emph{frame-based (or fidelity)}, individual time frames containing detected peaks in the explanation map are replaced with their reconstructions~\cite{nguyen2024robust}, independent of human labels. 
In \emph{segment-based}, entire 1-second segments are replaced if they contain at least one peak and overlap with human annotations to avoid dominating the reconstruction. This constrains the test to plausible anomaly regions, aligns with the concept of comprehensibility and human-aligned evaluation~\cite{chan2022comparative}~\cite{li2023survey}, and complements the F-score.

\begin{figure}[t!]
          \centering
          \includegraphics[width=0.95\linewidth]{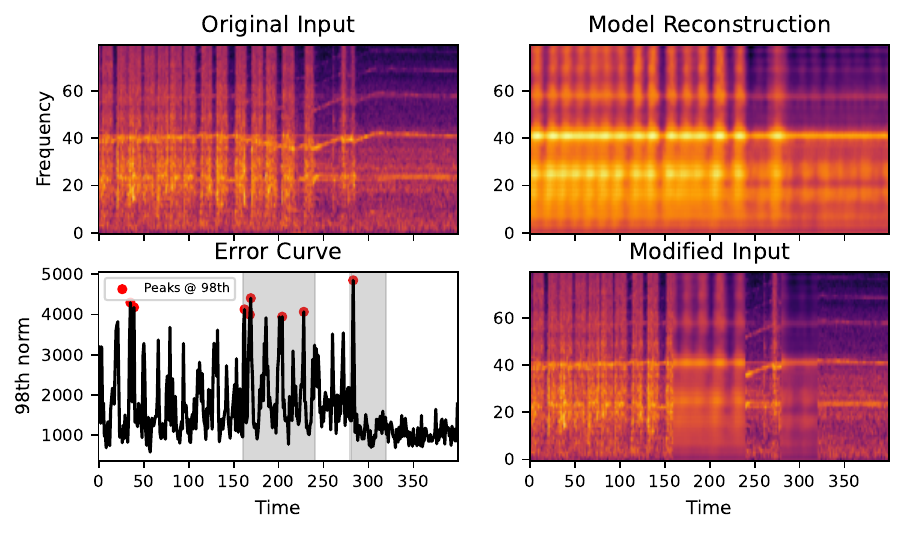}
          \caption{Segment-based input replacement. Gray areas indicate regions selected by the MAE error map at the 98th percentile.}
          \label{fig:seg_example}
        \end{figure}

Figure \ref{fig:seg_example} shows a concrete example of segment-based replacement using the 1D MAE error map, where masking the identified regions effectively reduces the broken anomaly signal, and simulates a more normal pattern.

\section{Experiments and Results}
\label{sec:exp}

We follow the same training setup as in~\cite{deschenes2025planing}, using AdamW~\cite{loshchilov2017decoupled} for 500 epochs with batch size 32. The learning rate starts at $10^{-3}$, scheduled with a cosine annealing schedule~\cite{loshchilov2017sgdr}, with a minimum of $10^{-5}$ and five warm-up cycles. Early stopping with patience of 30 epochs and model checkpointing based on validation loss are used. All experiments were run on a single 32GB GPU.

\textbf{Anomaly Detection:} The test AUC achieved by the standard autoencoder (AE) was 0.916, while the masked autoencoder (MAE) reached 0.902. However, due to the nondeterministic nature of the training, we repeated the experiments on five different random seeds. On average, AE achieved an AUC of $0.885 \pm 0.032$, while MAE achieved $0.864 \pm 0.048$. Despite this minor drop in detection performance, the MAE consistently produced more informative and localized attribution maps, as shown in interpretability evaluations.

\subsection{Interpretability Evaluation}

\begin{figure*}[t!]
    \centering
    \begin{subfigure}[b]{0.3\textwidth}
        \centering
        \includegraphics[width=0.9\linewidth]{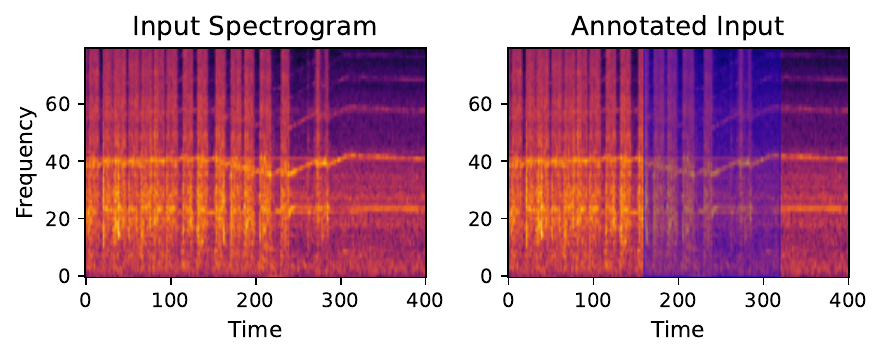}
        \caption{Input spectrogram}
        \label{fig:broken_inputs}
    \end{subfigure}
    \hfill
    \begin{subfigure}[b]{0.95\textwidth}
        \centering
        \includegraphics[width=\linewidth]{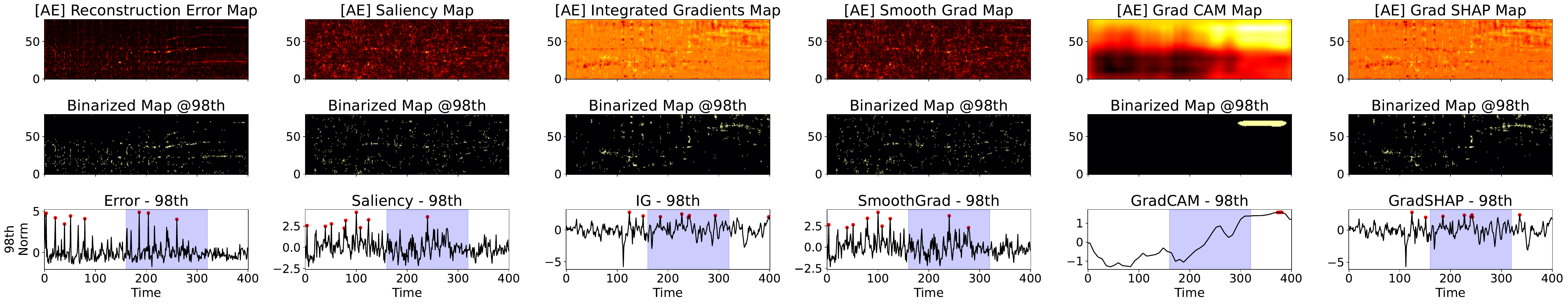}
        \caption{Attribution maps for the AE}
        \label{fig:ae_maps}
    \end{subfigure}
    \hfill
    \begin{subfigure}[b]{0.95\textwidth}
        \centering
        \includegraphics[width=\linewidth]{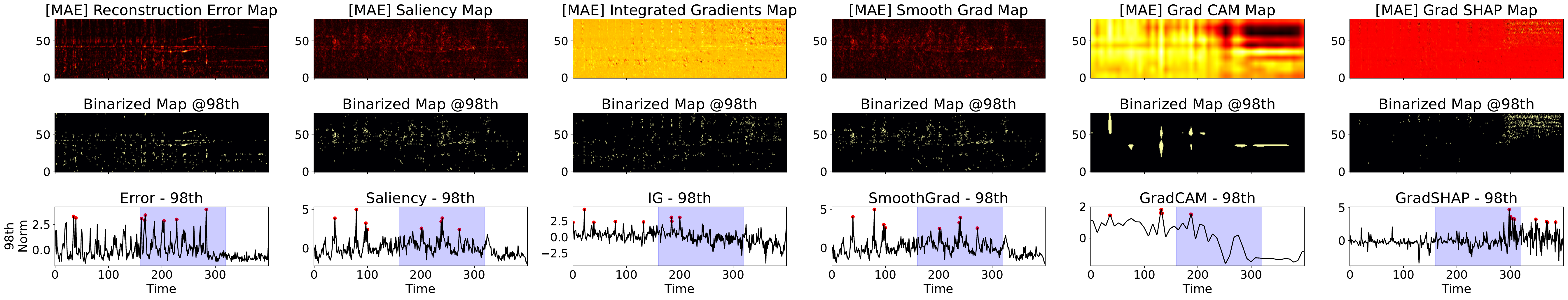}
        \caption{Attribution maps for the MAE}
        \label{fig:mae_maps}
    \end{subfigure}
    \caption{Qualitative comparison of interpretability methods on a broken board anomaly.
    (a) shows the input spectrogram, with the input overlaid by human-annotated anomaly intervals.
    (b) and (c) display the raw 2D attribution maps from six methods, following their corresponding binarized masks at the 98th percentile, and the last row presents 1D temporal attribution signals collapsed over frequency at the 98th percentile, with red dots marking detected peaks and shaded regions indicating annotated anomalies, for the AE and MAE, respectively.}
    \label{fig:qualitative_examples}
\end{figure*}

\subsubsection{F-score}
\begin{figure}[t!]
    \centering
    \includegraphics[width=\linewidth]{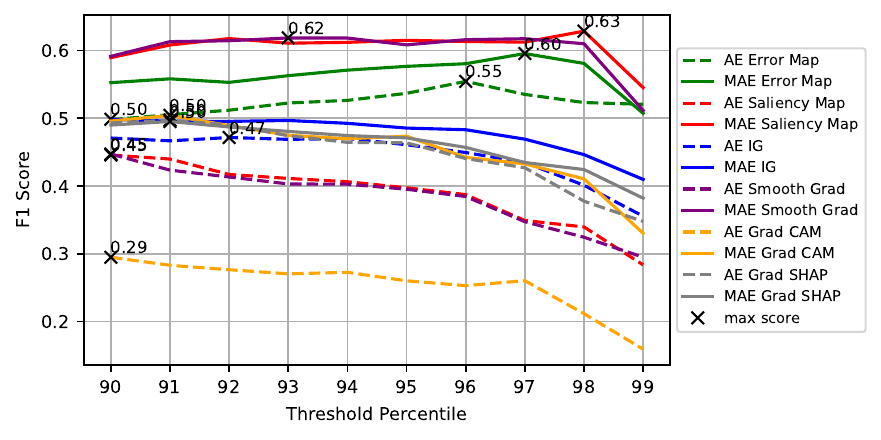}
    \caption{F-score across different percentile thresholds (90th–99th) for all interpretability methods applied to AE and MAE models. The highest score for each method is marked with an “x”.}
    \label{fig:fscore_plot}
\end{figure}

We evaluated six interpretability methods across a range of high percentile thresholds (90th–99th) using F-score as a quantitative proxy for anomaly localization quality. As shown in Figure~\ref{fig:fscore_plot}, the masked autoencoder (MAE) consistently outperforms the standard autoencoder (AE) across all methods. In particular, 
the highest score of 0.63 was achieved by MAE's saliency map at the 98th percentile, compared to AE's best of 0.55 from the error map at the 96th percentile.

\subsubsection{Faithfulness}
\begin{figure}[t!]
    \centering
    \includegraphics[width=\linewidth]{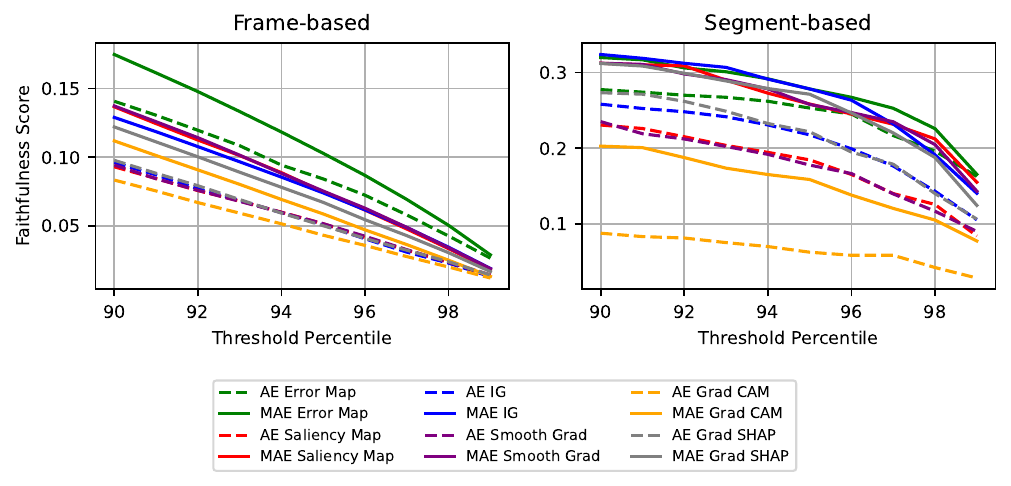}
    \caption{Faithfulness scores across percentile thresholds (90th–99th).}
    \label{fig:faithfulness_scores}
\end{figure}

We evaluate the faithfulness of each method by measuring the drop in reconstruction error after replacing detected regions with the model’s reconstruction.
From Figure~\ref{fig:faithfulness_scores}, MAE-based explanations consistently yield higher faithfulness scores than AE-based ones across all methods and thresholds. 
The MAE Error Map, in particular, stands out as the most faithful explanation, at higher percentiles (95–98\%), where it corresponds to the few most anomalous regions.

\subsubsection{Qualitative Analysis}
Figure~\ref{fig:qualitative_examples} presents visual comparisons between attribution maps for a broken board anomaly. The MAE error map produces more localized and structured attribution map; that outlines the normal straight and abnormal non-straight horizontal lines within the annotated anomaly region, and more peaks are detected inside the annotated region, aligning well with human annotations. In contrast, the other attribution maps often produce scattered attributions or focus on irrelevant regions.

In summary, the results show that the MAE, despite a slightly lower detection AUC than the AE, produces better anomaly detection explanations. Among the MAE-based methods, the error map is the most reliable across both F-score and faithfulness evaluations. Saliency Map and SmoothGrad provide strong event localization, while Integrated Gradients and GradSHAP yield more faithful but less precise attributions. GradCAM performs the worst overall. This suggests the MAE's potential for reliable explanations in real-world applications.
\section{Conclusion}
\label{sec:conclude}

In this work, we investigated the interpretability of autoencoder-based acoustic anomaly detection using real industrial data, comparing a standard autoencoder (AE) and a masked autoencoder (MAE). We evaluated six post-hoc explanation methods using plausibility and faithfulness metrics. Although MAE achieved a slightly lower detection AUC than the AE model, it produced more temporally precise and faithful attribution maps, highlighting its strength in generating meaningful explanations. Among methods, MAE error maps offered consistent good detection precision, faithfulness, and structured attribution map.
These findings suggest that MAE-based models improve interpretability with minimal impact on detection performance, making them suited for industrial applications where trust and actionable insights are essential.

\bibliographystyle{IEEEbib}
\bibliography{strings,refs}

@inproceedings{deschenes2025planing,
  title={Planing it by ear: Convolutional neural networks for acoustic anomaly detection in industrial wood planers},
  author={Desch{\^e}nes, A. and Georges, R. and Subakan, C. and Ugulino, B. and Henry, A. and Morin, M.},
  booktitle={Proc. of ICASSP},
  pages={1--5},
  year={2025},
  organization={IEEE}
}

@article{BUEHLMANN2002197,
title = {Impact of human error on lumber yield in rough mills},
journal = {Robotics and Computer-Integrated Manufacturing},
volume = {18},
number = {3},
pages = {197-203},
year = {2002},
note_ = {11th International Conference on Flexible Automation and Intelligent Manufacturing},
issn = {0736-5845},
doi = {https://doi.org/10.1016/S0736-5845(02)00010-8},
url = {https://www.sciencedirect.com/science/article/pii/S0736584502000108},
author = {U. Buehlmann and R. {E. Thomas}}
}

@article{li2023survey,
  title={A survey on explainable anomaly detection},
  author={Li, Z. and Zhu, Y. and Van Leeuwen, M.},
  journal={ACM Transactions on Knowledge Discovery from Data},
  volume={18},
  number={1},
  pages={1--54},
  year={2023},
  publisher={ACM New York, NY}
}

@article{vaswani2017attention,
  title={Attention is all you need},
  author={Vaswani, A. and Shazeer, N. and Parmar, N. and Uszkoreit, J. and Jones, L. and Gomez, A. N and others},
  journal={Advances in neural information processing systems},
  volume={30},
  year={2017}
}

@inproceedings{he2022masked,
  title={Masked autoencoders are scalable vision learners},
  author={He, K. and Chen, X. and Xie, S. and Li, Y. and Doll{\'a}r, P. and Girshick, R.},
  booktitle={Proc. of CVPR},
  pages={16000--16009},
  year={2022}
}

@inproceedings{kong2023understanding,
  title={Understanding masked autoencoders via hierarchical latent variable models},
  author={Kong, L. and Ma, M. Q and Chen, G. and Xing, E. P and Chi, Y. and Morency, L.-P. and Zhang, K.},
  booktitle={Proc.of CVPR},
  pages={7918--7928},
  year={2023}
}

@article{georgescu2023masked,
  title={Masked autoencoders for unsupervised anomaly detection in medical images},
  author={Georgescu, M.-I.},
  journal={Procedia Computer Science},
  volume={225},
  pages={969--978},
  year={2023},
  publisher={Elsevier}
}

@inproceedings{zimmerer2019unsupervised,
  title={Unsupervised anomaly localization using variational auto-encoders},
  author={Zimmerer, D. and Isensee, F. and Petersen, J. and Kohl, S. and Maier-Hein, K.},
  booktitle={Proc. of MICCAI},
  pages={289--297},
  year={2019},
  organization={Springer}
}

@article{simonyan2013deep,
  title={Deep inside convolutional networks: Visualising image classification models and saliency maps},
  author={Simonyan, K. and Vedaldi, A. and Zisserman, A.},
  journal={arXiv preprint arXiv:1312.6034},
  year={2013}
}

@inproceedings{sundararajan2017axiomatic,
  title={Axiomatic attribution for deep networks},
  author={Sundararajan, M. and Taly, A. and Yan, Q.},
  booktitle={Proc. of ICML},
  pages={3319--3328},
  year={2017},
  organization={PMLR}
}

@article{lundberg2017unified,
  title={A unified approach to interpreting model predictions},
  author={Lundberg, S. M and Lee, S.-I.},
  journal={Advances in neural information processing systems},
  volume={30},
  year={2017}
}

@article{smilkov2017smoothgrad,
  title={Smoothgrad: removing noise by adding noise},
  author={Smilkov, D. and Thorat, N. and Kim, B. and Vi{\'e}gas, F. and Wattenberg, M.},
  journal={arXiv preprint arXiv:1706.03825},
  year={2017}
}

@inproceedings{selvaraju2017grad,
  title={Grad-cam: Visual explanations from deep networks via gradient-based localization},
  author={Selvaraju, R. R and Cogswell, M. and Das, A. and Vedantam, R. and Parikh, D. and Batra, D.},
  booktitle={Proceedings of ICCV},
  pages={618--626},
  year={2017}
}

@inproceedings{el2023multivariate,
  title={Multivariate time series anomaly detection: Fancy algorithms and flawed evaluation methodology},
  author={El Amine Sehili, M. and Zhang, Z.},
  booktitle={Proc. of Tech. Conf. on Perf. Eval. and Benchmarking},
  pages={1--17},
  year={2023},
  organization={Springer}
}

@inproceedings{chan2022comparative,
  title={A Comparative Study of Faithfulness Metrics for Model Interpretability Methods},
  author={Chan, C. S. and Kong, H. and Guanqing, L.},
  booktitle={Proc. of ACL},
  pages={5029--5038},
  year={2022}
}

@article{loshchilov2017decoupled,
  title={Decoupled weight decay regularization},
  author={Loshchilov, I. and Hutter, F.},
  journal={arXiv preprint arXiv:1711.05101},
  year={2017}
}

@inproceedings{
loshchilov2017sgdr,
title={{SGDR}: Stochastic Gradient Descent with Warm Restarts},
author={I. Loshchilov and F. Hutter},
booktitle={Proc. of ICLR},
year={2017},
url={https://openreview.net/forum?id=Skq89Scxx}
}

@inproceedings{marchi2015novel,
  title={A novel approach for automatic acoustic novelty detection using a denoising autoencoder with bidirectional LSTM neural networks},
  author={Marchi, E. and Vesperini, F. and Eyben, F. and Squartini, S. and Schuller, B.},
  booktitle={Proc. of ICASSP},
  pages={1996--2000},
  year={2015},
  organization={IEEE}
}

@inproceedings{Koizumi2020DescriptionAD,
  title={Description and Discussion on DCASE2020 Challenge Task2: Unsupervised Anomalous Sound Detection for Machine Condition Monitoring},
  author={Y.Koizumi and Y. Kawaguchi and K. Imoto and T. Nakamura and Y. Nikaido and R. Tanabe and H. Purohit and others},
  booktitle={Workshop on Detection and Classification of Acoustic Scenes and Events},
  year={2020},
  url={https://api.semanticscholar.org/CorpusID:219559355}
}

@article{bouman2025autoencoders,
  title={Autoencoders for Anomaly Detection are Unreliable},
  author={Bouman, R. and Heskes, T.},
  journal={arXiv preprint arXiv:2501.13864},
  year={2025}
}

@article{antwarg2019explaining,
  title={Explaining anomalies detected by autoencoders using SHAP},
  author={Antwarg, L. and Miller, R. M. and Shapira, B. and Rokach, L.},
  journal={arXiv preprint arXiv:1903.02407},
  year={2019}
}

@article{roshan2021utilizing,
  title     = {Utilizing XAI technique to improve autoencoder based model for computer network anomaly detection with shapley additive explanation(SHAP)},
  author    = {K. Roshan and A. Zafar},
  journal   = {International Journal of Computer Networks \& Communications},
  year      = {2021},
  doi       = {10.5121/ijcnc.2021.13607},
  bibSource = {Semantic Scholar https://www.semanticscholar.org/paper/eb40fb0c53cda6a6a4709ba3b91162dad840d7a5}
}

@article{tritscher2023feature,
  title={Feature relevance XAI in anomaly detection: Reviewing approaches and challenges},
  author={Tritscher, J. and Krause, A. and Hotho, A.},
  journal={Frontiers in Artificial Intelligence},
  volume={6},
  pages={1099521},
  year={2023},
  publisher={Frontiers Media SA}
}

@article{huang2022self0supervised,
  title     = {Self-Supervised Masking for Unsupervised Anomaly Detection and Localization},
  author    = {Chaoqin Huang and Qinwei Xu and Yanfeng Wang and Yu Wang and Ya Zhang},
  journal   = {IEEE transactions on multimedia},
  year      = {2022},
  doi       = {10.1109/TMM.2022.3175611},
  bibSource = {Semantic Scholar https://www.semanticscholar.org/paper/1c0165247ce1d56a9de7be50ca6c4a49f0db4a82}
}

@article{lee2024explainable,
  title={Explainable time series anomaly detection using masked latent generative modeling},
  author={Lee, D. and Malacarne, S. and Aune, E.},
  journal={Pattern Recognition},
  volume={156},
  pages={110826},
  year={2024},
  publisher={Elsevier}
}

@article{holly2022autoencoder,
  title     = {Autoencoder based Anomaly Detection and Explained Fault Localization in Industrial Cooling Systems},
  author    = {S. Holly and R. Heel and D. Katic and L. Schoeffl and A. Stiftinger and P. Holzner and T. Kaufmann and others},
  journal   = {PHM Society European Conference},
  year      = {2022},
  doi       = {10.36001/phme.2022.v7i1.3349},
  bibSource = {Semantic Scholar https://www.semanticscholar.org/paper/3d07863c544b7ad426f9b7e11ba08107bb7ccf79}
}

@article{vsimic2025comprehensive,
  title={A comprehensive analysis of perturbation methods in explainable AI feature attribution validation for neural time series classifiers},
  author={{\v{S}}imi{\'c}, I. and Veas, Eduardo and Sabol, V.},
  journal={Scientific Reports},
  volume={15},
  number={1},
  pages={26607},
  year={2025},
  publisher={Nature Publishing Group UK London}
}

@inproceedings{gungor2024robust,
  title={A robust framework for evaluation of unsupervised time-series anomaly detection},
  author={Gungor, O. and Rios, A. and Mudgal, P. and Ahuja, N. and Rosing, T.},
  booktitle={Proc. of ICPR},
  pages={48--64},
  year={2024},
  organization={Springer}
}

@article{nguyen2024robust,
  title={Robust explainer recommendation for time series classification},
  author={Nguyen, T. T. and Le Nguyen, T. and Ifrim, G.},
  journal={Data Mining and Knowledge Discovery},
  volume={38},
  number={6},
  pages={3372--3413},
  year={2024},
  publisher={Springer}
}

@article{powers2020evaluation0,
  title   = {Evaluation: from precision, recall and F-measure to ROC, informedness, markedness and correlation},
  author  = {D. M. W. Powers},
  year    = {2020},
  journal = {arXiv preprint arXiv: 2010.16061}
}

@article{lapuschkin2019unmasking,
  title={Unmasking Clever Hans predictors and assessing what machines really learn},
  author={Lapuschkin, S. and W{\"a}ldchen, S. and Binder, A. and Montavon, G. and Samek, W. and M{\"u}ller, K.-R.},
  journal={Nature communications},
  volume={10},
  number={1},
  pages={1096},
  year={2019},
  publisher={Nature Publishing Group UK London}
}

\end{document}